\documentclass[%
 reprint,
superscriptaddress,
 amsmath,amssymb,
 aps,
]{revtex4-2} 
\usepackage{amsmath}
\usepackage[ruled,vlined]{algorithm2e}
\usepackage{amsfonts}
\usepackage{bm}
\usepackage{xcolor}
\usepackage{verbatim}
\usepackage{graphicx}
\usepackage{siunitx}
\usepackage{upgreek}
\usepackage[normalem]{ulem} 
\usepackage{ tipa }  
\usepackage{xr} 
\usepackage{gensymb} 
\usepackage{placeins}

\begin{document}

\raggedbottom
\author{Diksha Garg}
\email{d.garg@exeter.ac.uk}
\affiliation{Department of Physics and Astronomy, University of Exeter, Exeter, EX4 4QL, UK.}
\author{Anil Kumar}
\affiliation{School of Engineering, University of Warwick, Coventry, CV4 7AL, UK.}
\author{Ian R. Hooper}
\affiliation{Department of Physics and Astronomy, University of Exeter, Exeter, EX4 4QL, UK.}
\author{Nicholas E.~Grant}
\affiliation{School of Engineering, University of Warwick, Coventry, CV4 7AL, UK.}
\author{John D.~Murphy}
\affiliation{School of Engineering, University of Warwick, Coventry, CV4 7AL, UK.}
\affiliation{School of Engineering, University of Birmingham, Edgbaston, Birmingham, B15 2TT, UK.}
\author{John E.~Cunningham}
\affiliation{School of Electronic and Electrical Engineering, University of Leeds, Leeds, LS2 9JT, UK.}
\author{Nick Stone}
\affiliation{Department of Physics and Astronomy, University of Exeter, Exeter, EX4 4QL, UK.}
\author{Euan Hendry}
\affiliation{Department of Physics and Astronomy, University of Exeter, Exeter, EX4 4QL, UK.}
\author{Harry Penketh}
\email{h.penketh2@exeter.ac.uk}
\affiliation{Department of Physics and Astronomy, University of Exeter, Exeter, EX4 4QL, UK.}

\title{Real-time THz ptychography using an optically modulated aperture}

\begin{abstract}
Ptychography allows ordinary cameras to perform both quantitative phase and intensity imaging. However, the requirement for relative object to probe motion and high measurement count significantly limits imaging speed and application scope. In this work, we introduce optically modulated aperture ptychography, using a silicon photomodulator as a rapidly reconfigurable aperture, deployable across the microwave, millimeter wave and terahertz bands. We demonstrate experimentally at $\sim$ 0.1~THz real-time ptychographic image capture, without moving parts. Furthermore, a reconfigurable aperture facilitates diverse and spatially multiplexed probes. We use a multiplexed known-probe reconstruction strategy that improves signal-to-noise ratio and increases the information content per measurement, enabling the capture of fluid dynamics at 16 fps. Our optical modulation approach enables high-throughput ptychographic imaging of dynamic samples, for applications such as non-destructive evaluation and in-vivo biomedical imaging.

\end{abstract}
\maketitle


\section{Introduction}
Standard cameras typically capture only the intensity of an electromagnetic field and therefore record incomplete information about a scene. Techniques such as digital holography \cite{latychevskaia2019iterative} and coherent diffractive imaging \cite{zhang2016phase} provide the means to retrieve the missing phase component, enabling lensless imaging \cite{boominathan2021recent}, aberration removal \cite{ferraro2008full}, and high contrast imaging of dielectric samples \cite{utadiya2023digital}. Ptychography is a leading coherent diffractive imaging method, demonstrated across several spectral bands \cite{wang2025ptychography}. When compared with holography it offers a much simpler experimental implementation without a reference beam, at the expense of imaging speed \cite{monaco2022comparison, maiden2015quantitative, rodenburg2007transmission}. 

In conventional scanning ptychography, object phase is reconstructed from diffraction patterns recorded at overlapping illumination positions in the object plane \cite{rodenburg2019ptychography}. A physical aperture is often used to localize the illumination beam, while the object is moved to generate the overlapping positions. Fly-scanning ptychography is a variant in which continuous motion of the object is used, reducing delays caused by object acceleration and deceleration between measurements \cite{huang2015fly}. However,
the relative movement of the object across the illumination still imposes a mechanical restriction on imaging speeds, as hundreds to thousands of diffraction patterns are typically needed.

Several multiplexed aperture techniques have been demonstrated which reduce the number of diffraction patterns required for reconstruction \cite{lin2021parallel, he2018high, aastrand2024adaptive, lyubomirskiy2022multi, sidorenko2015single}. However, these approaches still require object scanning and are sensitive to artifacts due to the interference between apertures. Another popular ptychographic approach, Fourier ptychography, addresses these limitations to some extent by replacing object motion with a changing illumination angle \cite{zheng2021concept, konda2020fourier}. Indeed, real-time Fourier ptychography has been demonstrated in the visible spectral region by using reconfigurable LED arrays to rapidly scan illumination angle \cite{zheng2021concept, konda2020fourier,tian2015computational}. 

Whilst microscopy at visible wavelengths has benefited substantially from ptychographic augmentation, ptychography in other frequency bands is underdeveloped. In particular, the problematic mm-wave and terahertz (THz) bands, in-between the more developed infrared and microwave spectral regions, could benefit enormously from the unrealized potential in ptychographic imaging. Due to the transparency of most non-polar dielectrics in the THz region, it has great promise for label-free, safe and non-destructive imaging across a wide range of applications \cite{li2018study}. The high sensitivity of THz radiation to material hydration also makes it a promising tool for differentiating normal from cancerous tissues in biomedical imaging \cite{taylor2011thz}.

Simple demonstrations of THz ptychography have been realized \cite{valzania2018terahertz, mukherjee2026terahertz}, and the straightforward implementation when compared to two-beam holographic systems \cite{Heimbeck2020THzHolography,Penketh2024Holography} may be a key facilitator of translation from the laboratory to diverse end-users. Real-time THz ptychography could enable, for example, high throughput non-destructive evaluation, monitoring of concealed fluid dynamics and in-vivo biological observation. However, there is no available analog to a reconfigurable LED array in this frequency range, and imaging times remain a central challenge. Thus, despite many prospective applications, real-time ptychography has not been demonstrated in this spectral region.
In this work, we overcome the limitations of scanning ptychography imposed by the requirement for physical movement by utilizing an optically modulated spatial light modulator (SLM). The proposed SLM-based approach enables localized modulation of THz transmission through a silicon wafer and forms an optically controlled THz aperture between photoexcited and non-excited regions. In this configuration, the overlapping scan positions can be realized by changing the optical illumination. Rapidly scanning the aperture, in place of moving the object to be imaged, yields a step change in accessible imaging speed applicable across the mm-wave and THz bands. Furthermore, such a reconfigurable aperture opens up a swathe of opportunities for spatial multiplexing within the probe function, which reduces the number of necessary diffraction measurements and further improves imaging speed. Leveraging this, we demonstrate real-time ptychographic imaging, capturing fluid dynamics in a sample at 16~fps.

\begin{figure*}[tp!]
\centering
\includegraphics[width=0.8\textwidth]{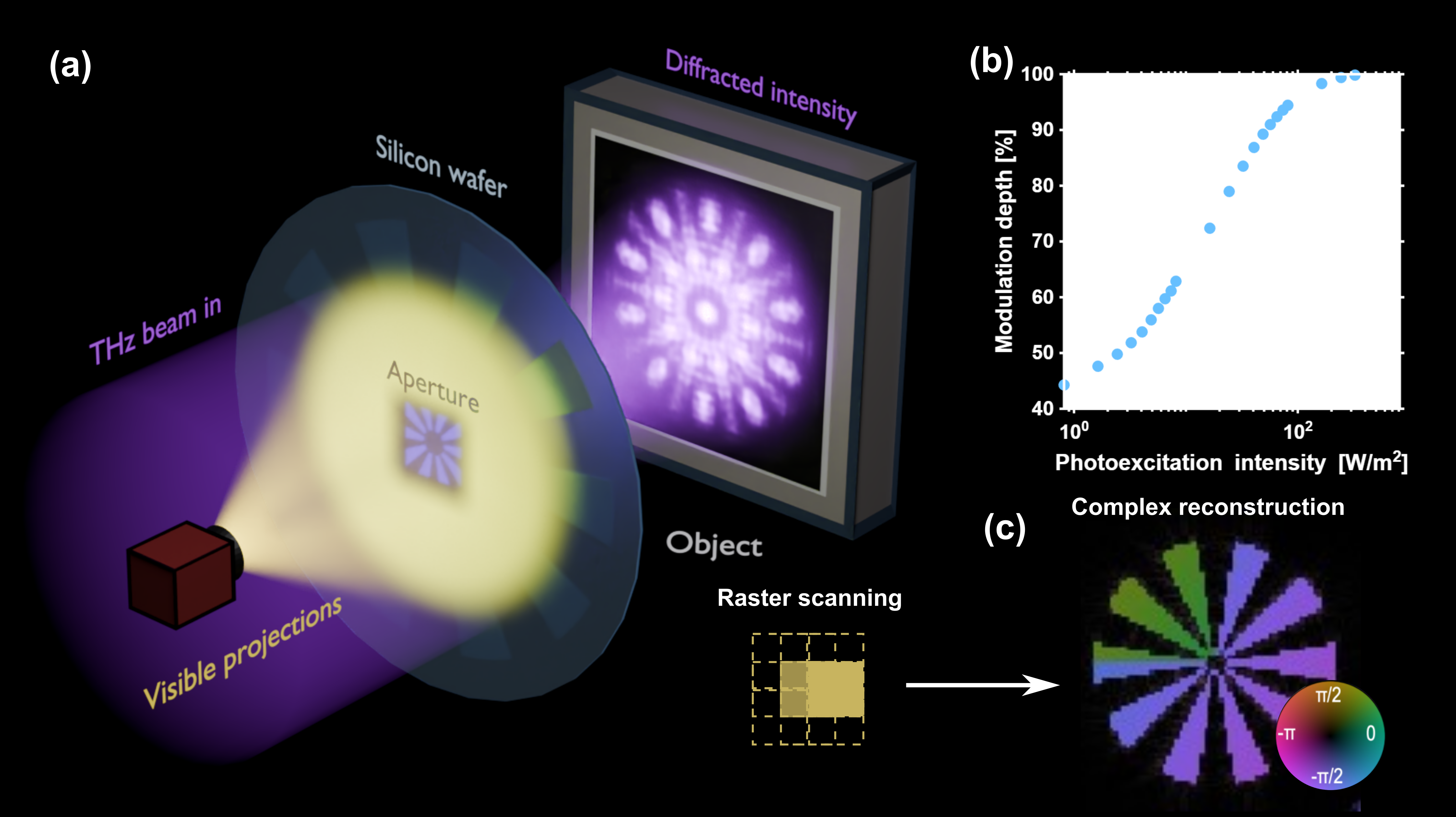} 
\caption{Principle of optically modulated
aperture ptychography. (a) Spatially structured optical illumination is projected onto a silicon wafer to form a transient aperture for the incident THz beam. The transmitted THz radiation interacts with the object placed behind the wafer, producing a diffraction pattern from the localized region. (b) Measured reduction in the THz transmission (modulation depth) through a white light illuminated 380~$\upmu$m thick passivated silicon wafer at varying illumination intensities. (c) Numerically simulated reconstruction from a sequence of aperture positions, forming a complex-valued object with uniform intensity (brightness) and spatially varying phase (color) as shown in the inset.}
\label{FIG:Setup}
\end{figure*}

\section{Method}
\subsection{Conventional scanning ptychography}
In standard scanning ptychography, the illumination is shaped into a spatially localized probe $P(r)$, typically via beam focusing or with a static aperture. This confinement ensures that the measured diffraction intensity varies with the relative motion between the probe and object $O(r)$. In the optical regime, it is common for the probe to remain fixed whilst the object is translated to scan position $\mathbf{r}_n$, yet it is conventional to describe the exit wave from the probed object as \cite{rodenburg2004phase}
\begin{equation}
     \psi_n(\mathbf{r}) = P(\mathbf{r} - \mathbf{r}_n)\, O(\mathbf{r})
\end{equation}
and the measured intensity at a downstream camera as
\begin{equation}
    I_n(\mathbf{q}) = |\mathcal{P}_z \left\{ \psi_n(\mathbf{r}) \right\}|^2,
\end{equation}
where $\mathcal{P}_z$ is the diffraction propagator between object and detector planes, with the latter described by position vector $\mathbf{q}$. Unlike in optical ptychography where $\mathcal{P}_z$ would usually be the Fourier transform, at THz frequencies the detector typically sits in the Fresnel diffraction zone, which can be modeled using the angular spectrum method \cite{poon2014introduction}.
Eqs. 1 and 2 describe the forward model for ptychography.

For a sequence of $N$ overlapping scan positions such that each point on the object is probed at least twice, the unknown complex valued object can be reconstructed iteratively using established ptychographic solvers such as ePIE \cite{maiden2009improved} and RAAR \cite{enfedaque2019high}. If the probe overlap is greater, the additional redundancy facilitates robust retrieval of both the object $O(r)$ and unknown probe $P(r)$, in so called blind ptychography. See ref. \cite{zhang2021analysis,mei2024iterative} for an in depth discussion of phase retrieval algorithms for ptychography.

\subsection{Optically modulated aperture ptychography}
The critical innovation of this work is the application of a dynamic photo-reconfigurable aperture created using a silicon photomodulator. The photomodulator is illuminated with spatially patterned visible light, which modifies the local conductivity of the photomodulator and consequently, the transmission of the THz beam passing through it. Regions exposed to visible light strongly attenuate THz transmission, while non-illuminated areas allow the THz radiation to pass, thereby creating a structured probe. To induce sufficient photoconductivity to render photo-excited regions opaque, it is necessary to passivate the surfaces of the silicon wafer, a process that increases the effective lifetime of photogenerated charge carriers, which can be tuned from $\upmu$s to ms (see experimental details section D) \cite{hooper2019high}. Passivation reduces the required visible photoexcitation intensity and enables the use of safe, low-cost continuous wave visible light sources such as LEDs. For this study, we choose two different modulators with carrier lifetimes of $\sim$ 4~ms and $\sim$ 1~ms (Supplementary Fig. S2), the latter permitting a faster measurement rate.

 Fig.~\ref{FIG:Setup} (a) illustrates the concept, with the example of imaging a Siemens star located behind the photomodulator. The incident THz beam is expanded to match the modulation area, which is given by the extent of the illumination by visible light. By introducing a transparent aperture (a region without illumination) we can define a local THz probe region, as shown in Fig.~\ref{FIG:Setup} (a). The position of the aperture may be optically updated to new probe positions $\mathbf{r}_n$ using a digital micromirror device (DMD), which is capable of running at kHz rates. Here we tune the rate at which we update the aperture position to the lifetime of the modulator and the camera frame rate. As there are no moving parts in this implementation and the object is static, the imaging field of view is limited by the detector size and distance. Mimicking the traditional scanning ptychography approach, we raster scan a transparent square within our photomodulator (of size 26~mm $\times$ 26~mm) across the modulation area in a regular 2D grid of 9 mm step size. This leads to a linear overlap ratio of 65 percent. In results section 3B,  we explore the advantages of alternative probing strategies.

\subsection{Known-probe reconstruction algorithm}
As our probe function $P(r)$ can no longer be considered fixed for all measurements, we adapt the standard ptychographic formulation for the exit wave to
\begin{equation}
     \psi_n(\mathbf{r}) = P_n(\mathbf{r})\, O(\mathbf{r}).
     \label{eq:knownProbe}
\end{equation}
This allows us to use a standard ptychographic iterative algorithm, in this case PIE \cite{sindhya2011preference}, in which we only solve for the unknown object $O(\mathbf{r})$. The probe is treated as a known input, which in our case updates according to the optical projection sequence and photomodulator carrier dynamics. 

In practice, due to imperfect passivation, the carrier lifetime will vary slightly across the modulator (see Supplementary Fig. S1). Thus, we favour a reconstruction approach in which the probe spatio-temporal dynamics are first measured experimentally and then used to define a dynamic probe during reconstruction. The probe sequence $P_n(\mathbf{r})$ is measured by removing the object and placing the camera immediately behind the photomodulator. Since this measurement only constrains the intensity of the probe sequence, we assume a spatially uniform phase profile across the modulator plane ($\mathbf{r}$). In practice, this imparts local phase curvature, derived from the divergence of the THz beam, into the object response $O(\mathbf{r})$, which can then be removed by a one time no-object normalization. With this assumption, the probe intensity sequence and 
a random object guess are input into an adapted PIE algorithm \cite{Andrew_Maiden_Ptychography_algorithms_from}  following the Eq.~\ref{eq:knownProbe} formulation.
We use 100 iterations for image reconstruction, and 300 iterations for later video reconstructions in Fig.~\ref{FIG:video}.

\subsection{Experimental details}
We perform our experimental demonstration with a continuous wave, 95 GHz, 180 mW IMPATT-diode source (TeraSense). This source flood-illuminates our silicon photomodulator, 5~mm behind which is placed the object to be imaged. Our detector, a Tera-4096 $64~\times~64$ pixel, 1.5 mm pitch array formed from GaAs high mobility heterostructures (TeraSense) is positioned 50~mm behind the object and along the optical axis. An annotated photograph of the experimental system is provided in the Supplementary Fig. S3.  

To generate the reconfigurable THz aperture, patterned visible light is encoded with a DLP7000 (Vialux) DMD, which has a maximum refresh rate exceeding 20~kHz. White light illumination is provided by an SLS605 liquid light guide coupled xenon arc lamp (Thorlabs). The THz and visible beams are combined on axis by reflecting the THz off a 1.2 mm thick, 45$\degree$ quartz beamsplitter, which is optically transparent. The illumination intensity at the photomodulator is 200 W/m$^2$ for imaging. Later, we increase this illumination intensity to 490 W/m$^2$ for video-rate capture.

We use two photomodulators in this work. The first, used for imaging of static objects, is a 150~mm diameter, high resistivity ($>$7000~$\Omega$cm) silicon wafer that has undergone surface passivation through Al$_2$O$_3$ deposition following Ref. \cite{grant2024activation}, which gives a carrier lifetime of $\sim$ 4~ms at 200 W/m$^2$. For this modulator, 200 W/m$^2$ illumination results in near 100\% modulation of transmission, as shown in Fig. \ref{FIG:Setup} (b). The thickness of this modulator, 380~$\upmu$m, is chosen from available stock to give a Fabry-Perot resonance, with increased transmission, near our operating frequency (78 GHz and 95 GHz, respectively). Later, for the high-speed imaging, we use a different modulator ($>$10,000 $\Omega$cm, 100 mm diameter and 100 $\upmu$m thickness) with a shorter lifetime of $\sim$ 1~ms to allow video-rate capture. Further details on the photomodulator fabrication and characterization can be found in Supplementary Section 1.

\section{Results and discussion}
\subsection{Proof of concept}
\begin{figure}[htbp]
\centering
\includegraphics[width=0.45\textwidth]{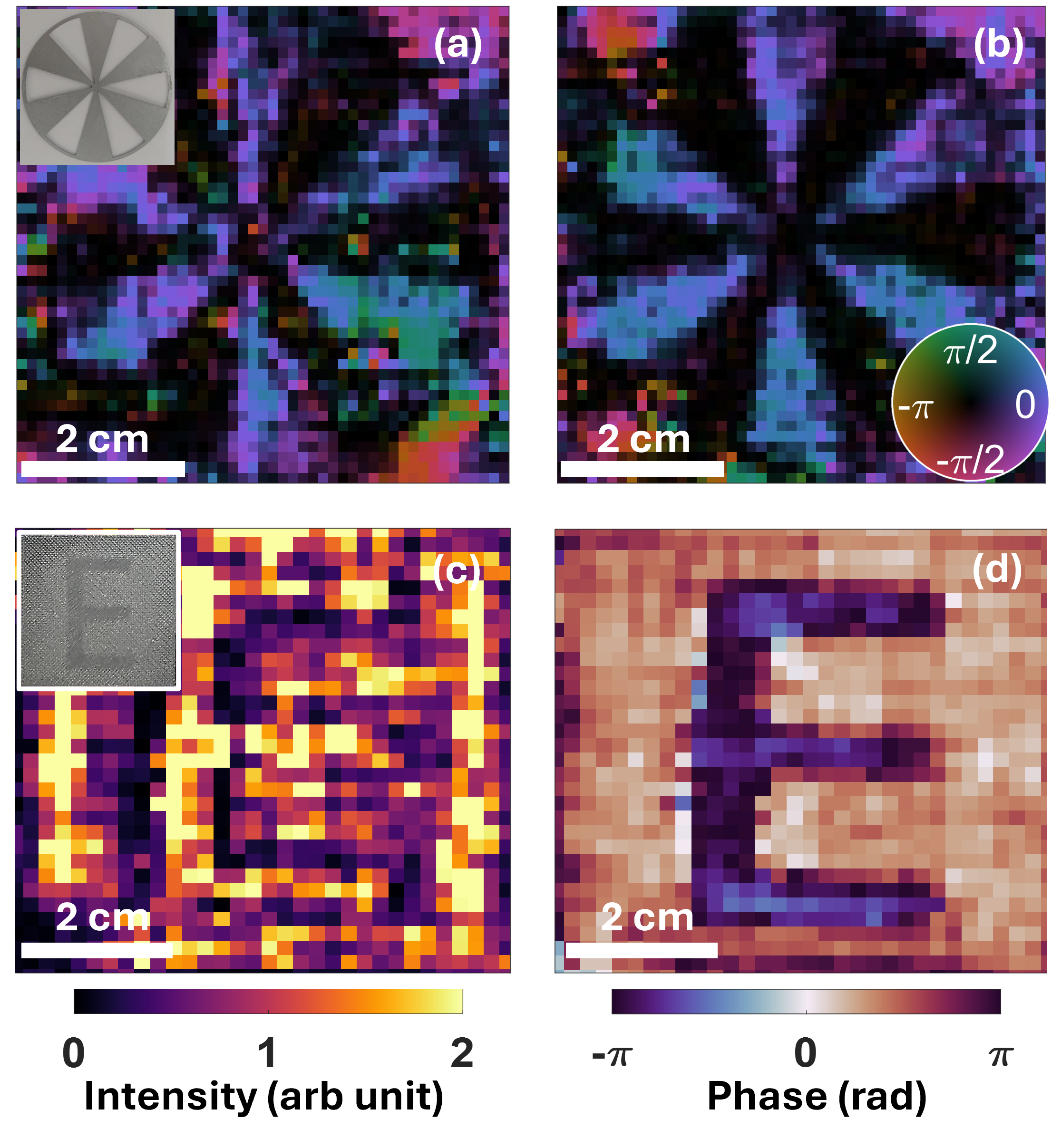} 
\caption{Complex reconstructed images of the Siemens star generated using optically modulated (a) raster, and (b) Hadamard probes. In these complex images, the colors represent phase values ranging from -$\uppi$ to $\uppi$, while the brightness indicates the relative intensity. (c) and (d) Intensity and phase images of the 3D-printed dielectric object generated using the Hadamard basis, the colormaps are from Ref. \cite{matplotlib_colormaps_matlab}. }
\label{FIG:basis}
\end{figure}

Exemplar imaging using an optically modulated aperture is shown in Fig. \ref{FIG:basis} (a). This is a complex-valued image of a metallic Siemens star, with colors representing phase and brightness indicating the local field intensity. For this we use a probe sequence of 8$~\times$~8 overlapping raster apertures, acquired with a camera frame rate of 45~fps. For each optical pattern, 12 continuous frames are recorded over a time interval of 267 ms. For reconstruction we use every 12th frame to ensure that carriers generated by the previous pattern do not contribute to measurement of the current pattern. This produces a $64~\times~64~\times~64$ pixel dataset or ptychogram, from 64 diffraction patterns recorded with a $64~\times~64~$ pixels camera in 17 seconds. Note that the measurement rate is slower than the minimum required by the modulator lifetime of $\sim$ 4~ms; data could therefore be acquired at a higher rate. The image in Fig. \ref{FIG:basis} (a) is reconstructed and normalized with a non-object reference (see Supplementary Fig. S4 for unnormalized reconstruction). Whilst the resulting image is clearly identifiable as the Siemens star, the signal-to-noise ratio (SNR) is low. We attribute this in part to the non-uniform spatial response of our THz camera \cite{mrnka2024rapid} and blinking hot‑pixel clusters at low signal counts (see Supplementary Fig. S5).

\subsection{Probe multiplexing}
To improve the SNR and imaging speed further, we leverage another key innovation afforded by our approach: a reconfigurable aperture offers new possibilities for spatial multiplexing. Typically in ptychography, a spatially multiplexed or multi probe approach requires each aperture to be in some way separable in the diffracted intensity, e.g. by position, wavelength or phase encoding \cite{yang2025optimizing, lyubomirskiy2022multi}. This requirement for separability limits the increase in throughput to a modest enhancement, following the number of apertures \cite{yao2020multi, hirose2020multibeam}. In this work, however, we find that by using a known-probe sequence and therefore reducing the number of unknowns to just the object $O(\mathbf{r})$, this constraint is relaxed. 

In keeping with the known-probe sequence strategy (Eq.~\ref{eq:knownProbe}), there is no requirement for separability between multiplexed apertures within a single projection. Therefore, we select our new probe basis to satisfy two criteria: 50$\%$ transparency across the entire modulation area, to maximize signal and throughput, and orthogonality between projected probes $P_n(\mathbf{r})$ to encourage even object sampling. This leads us naturally to the Hadamard basis. Unlike for the raster basis in which the individual aperture size must be larger than the sampling grid spacing, in this case the overlap is provided by the multiplexed nature of the Hadamard basis, and each point on the object is now probed N/2 times (N is the number of projections). To generate projectable binary optical patterns, the +1 and -1 entries in the Hadamard matrix are mapped to 255 and 0 values to DMD pixels, defining illuminated and non-illuminated apertures, respectively.

In Fig. ~\ref{FIG:basis} (b) we show that using the same acquisition parameters as in (a) but with an $8~\times~8$ Hadamard projection basis, we obtain a notable SNR improvement. With improved SNR, the resolution of our approach can be evaluated using the Siemens star spoke separation. We determine an object-plane resolution of 2.1$\pm$0.3~mm by identifying the radius at which the target spokes meet the Rayleigh criterion in field intensity (see Supplementary Fig. S6). Additionally, we note that due to spatial multiplexing in the Hadamard case, the number of iterations required for reconstruction may be significantly reduced (see Supplementary Fig. S7).

The phase information retrieved from scanning ptychography is essential for propagating the diffraction measurements to the object plane. Therefore, even an intensity-only object, as in Fig. ~\ref{FIG:basis} (b), requires accurate phase recovery to enable lensless reconstruction. However, the additional object phase images recovered have far broader utility, particularly for evaluating dielectric objects. In Fig. ~\ref{FIG:basis} (c) we show a normalized intensity-only image of such a dielectric object - a 3D-printed ABS 50 mm$\times$ 50 mm$\times$ 5~mm  block shown in the inset. Fig. ~\ref{FIG:basis} (d) is the corresponding phase-only image, in which small differences in optical path length manifest clear changes in measured phase, revealing a hidden inclusion with higher index in the form of the letter E. By inverting the relationship between phase delay and refractive index, the measured phase can be converted into a map of effective refractive index, shown in Supplementary Fig. S8. In S8, we show that the solved refractive distribution follows the known variation in infill density from 20\% to 100\% for the surround and letter E, respectively. 

\subsection{Video-rate imaging}
\begin{figure}[tp!]
\centering
\includegraphics[width=0.45\textwidth]{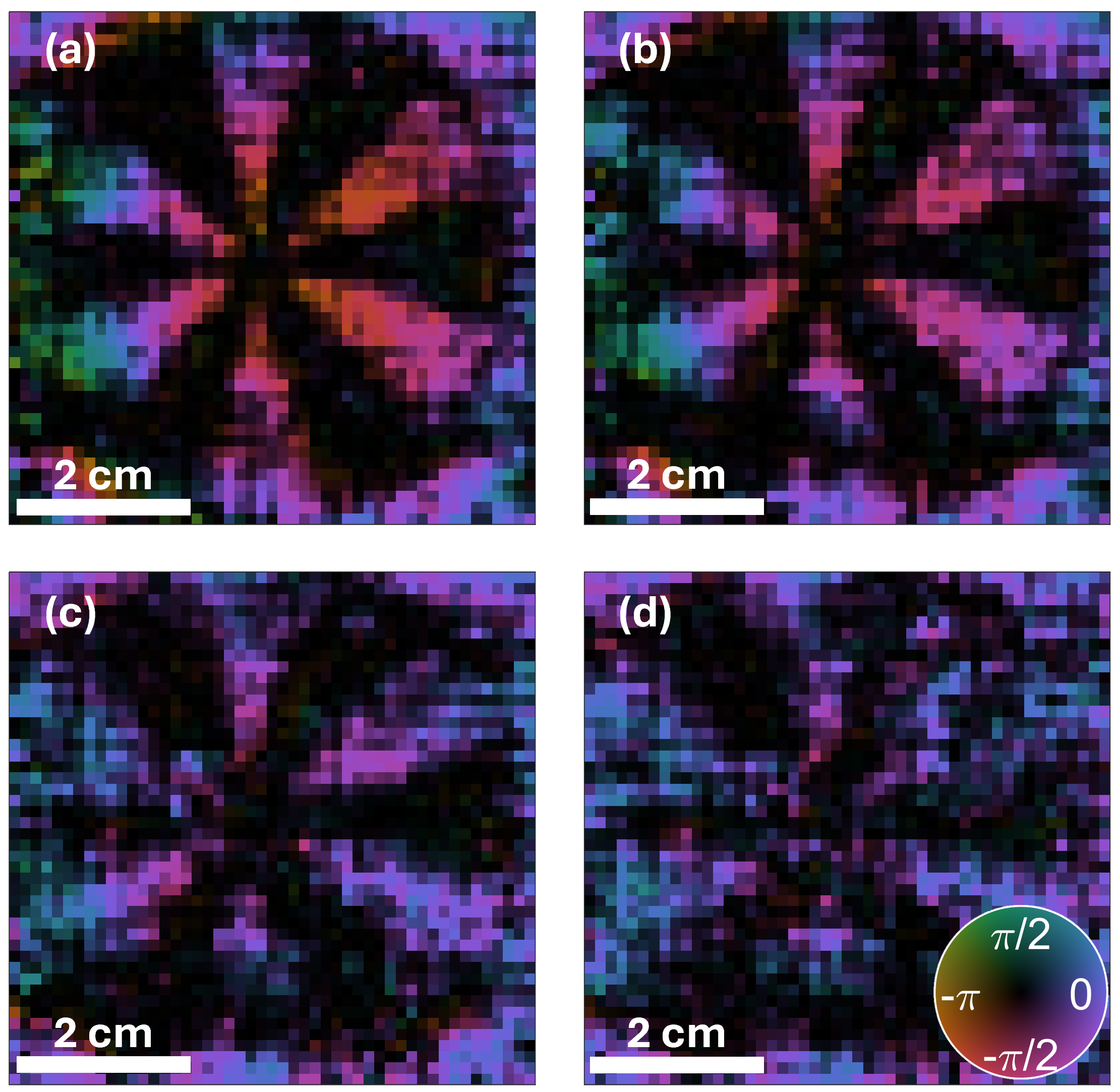} 
\caption{Complex images of the Siemens star reconstructed using varying number of random binary patterns: (a) 64, (b) 32, (c) 8 and (d) 4; in these complex images, the colors represent phase values ranging from -$\uppi$ to $\uppi$, while the brightness indicates the relative intensity. }
\label{FIG:undersampling}
\end{figure}

The optically modulated aperture ptychography demonstrated in Fig.~\ref{FIG:basis} achieves at least an order of magnitude reduction in measurement time relative to state-of-the-art THz scanning ptychography \cite{valzania2018terahertz}. Extending this improvement to video-rate acquisition would unlock the study of previously inaccessible dynamic phenomena and rapidly evolving samples. In this section we introduce our final innovation, in which we exploit the enhanced redundancy present in our spatially multiplexed ptychograms to increase imaging speeds further. 

\begin{figure*}[tp!]
\centering
\includegraphics[width=1\textwidth]{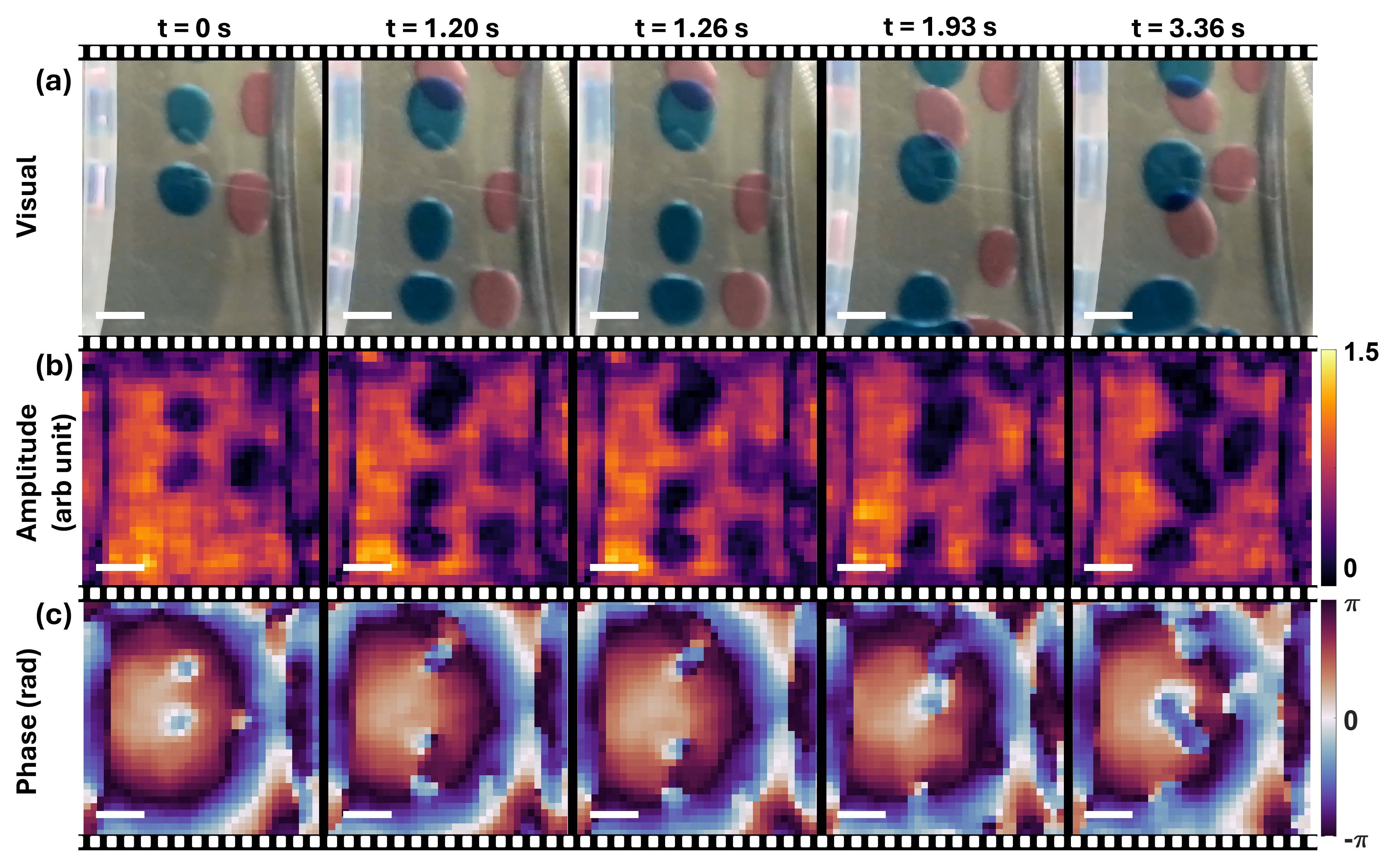} 
\caption{Visualization of downward motion of the water droplets through an oil medium. (a) Top row, visual images of the water-oil dynamic object at different times; the droplets are dyed in different colors to enhance visual contrast. Note that the visual images are acquired off axis from the THz system. (b) Middle and (c) bottom row, corresponding reconstructed amplitude and phase images. The scale bar in all images represents 1 cm. See Supplementary Visualization 1 for the continuous droplet motion over a longer time interval.}
\label{FIG:video}
\end{figure*}
 
In conventional scanning ptychography, each point on the object must be sampled redundantly to enable stable phase reconstruction. In our multiplexed scheme, complete object coverage can be achieved using as few as two binary projections. 
We investigate the effect of undersampling our measurement basis by using fewer measurements than are required for the raster-scanned case. To do this, we first substitute our Hadamard projection basis for random binary projections (in which 50$\%$ of the apertures are randomly open within each pattern). This removes any effects from the variation in spatial frequency content from pattern to pattern. 

Fig.~\ref{FIG:undersampling} examines the effect of undersampling the projected patterns formed using an $8~\times~8$ grid of such randomly distributed apertures. To minimize image noise for undersampled reconstructions, these figures are presented without normalization and therefore the phase curvature of the THz probe is visible. The fully sampled reconstruction, Fig.~\ref{FIG:undersampling} (a) is comparable to the Hadamard case in Fig.~\ref{FIG:basis} (b), despite the lack of basis orthogonality. 
In Fig.~\ref{FIG:undersampling} (b) the acquisition time is halved, with a modest reduction in the fidelity of the intensity reconstruction. As the number of optical patterns is reduced to 8 and 4 in (c) and (d) respectively, both intensity and phase fidelity continue to decrease, with phase diminishing more rapidly. In (d), corresponding to just four projected patterns, the object intensity response is recognizable but of heavily compromised quality.

To achieve video-rate imaging, we first employ a photomodulator with a shorter carrier lifetime ($\sim$ 1~ms), enabling photoexcited masks to dissipate faster between projections (see Supplementary Fig. S2 for wafer comparison). The illumination intensity is increased to  490 W/m$^2$ to compensate for the reduced carrier density associated with the shorter carrier lifetime. Under these conditions, the optical patterns are projected for a duration of 2 camera frames (a minimum imposed by the camera output trigger). At an increased camera frame rate of 67~fps, this corresponds to 30~ms per pattern. 

To improve reconstruction quality under highly undersampled conditions, the spatiotemporal correlations present in many real-world imaging applications can be exploited. In our final demonstration in Fig.~\ref{FIG:video}, we capture the motion of water droplets dispersed in mineral oil at a camera acquisition rate of 67 fps. 

To achieve high reconstruction frame rates with acceptable fidelity, we first generate a high fidelity reconstruction of the static scene using 300 frames. We then enter to a dynamic reconstruction mode, in which we update using only the 2 most recent projections, with the previous reconstruction supplied as the algorithm object guess. In this way, we ensure the scene is quasi-static over the reconstruction window (4 camera frames) and exploit the similarity of subsequent reconstructions to obtain a much higher fidelity image than that shown in Fig.~\ref{FIG:undersampling} (d), from fewer measurements (see Supplementary Table S1 for a summary of imaging times for each figure).

Supplementary Visualization 1 shows a real-time playback of the motion of colored water droplets dispersed in oil. In Fig.~\ref{FIG:video}, representative frames from Supplementary Visualization 1 are shown. For improved visualization, we display the field amplitude rather than the intensity and a $3~\times~3~\times~3$ spatiotemporal median filter is applied to the image sequence shown in Supplementary Visualization 1 (see Supplementary Visualization 2 for the unfiltered version).
The blue and pink dyes in the water drops are for visual distinction only, and correspond to two isolated parallel cavities separated by $\sim$ 3~mm along the optical axis (small enough to be inconsequential here). The slightly concave edges of the plastic containment vessel are visible left and right of the frames in all the data. Note that the THz reconstructions in this dataset are presented unnormalised to improve SNR. As such, the quadratic phase curvature of the illuminating beam can also be seen in the phase reconstructions.

In Fig.~\ref{FIG:video}, the water droplets are identified by significant reduction in transmission and a phase change relative to the oil background. The downward motion of the water droplets can be clearly tracked throughout the sequence, including the subtle displacement observed between reconstructed frames separated by only 60 ms, corresponding to an effective reconstructed frame rate exceeding 16 fps. This demonstration highlights the potential for detecting and tracking the presence of unwanted contaminants such as air bubbles and immiscible fluids in concealed pipes under flow conditions, as just one example of the many use cases unlocked by real-time THz ptychography.
\section{Conclusion}  
We have demonstrated a novel approach to THz ptychographic imaging using an optically reconfigured aperture. Using this system-level innovation we forgo the need for physical movement between the object and aperture and demonstrate real-time complex field imaging. Our silicon-based photomodulation approach is suitable for deployment across the millimeter-wave and THz bands. Furthermore, we exploit the massively parallel nature of optical control to generate spatially multiplexed apertures for increased SNR and imaging throughput, which we combine with temporal redundancy of the scene to capture fluid dynamics at 16 fps. It should be noted that the achieved imaging speed is constrained by the camera acquisition speed and SNR, and not the silicon photomodulator. This imaging framework overcomes the major limitation of mechanically scanned ptychography and expands the scope of millimeter-wave and THz imaging toward high-throughput, non-destructive evaluation of dielectric structures, as well as quantitative phase imaging for biomedical applications.

\section*{Funding}
 D.G, A.K, J.D.M., J.E.C., N.S., E.H. and H.P. acknowledge support from the Engineering and Physical Sciences Research Council via the Terabotics Programme Grant (EP/V047914/1). EH and I.R.H. also acknowledge support via the META4D Programme Grant (EP/Y015673/1), and J.D.M via (EP/S036261/1).

\section*{Disclosures}
The authors declare no conflicts of interest.

\section*{Data Availability} 
The research data underlying this work will be made openly available upon publication of the peer‑reviewed article.
 
\vspace{20cm}
\clearpage
\newpage
\newpage
\appendix
\setcounter{figure}{0}
\renewcommand{\thefigure}{S\arabic{figure}}
\setcounter{table}{0}
\renewcommand{\thetable}{S\arabic{table}}
\section*{Supplementary Material}
This supplementary document provides additional information supporting the main manuscript,``Real-time THz ptychography using an optically modulated aperture''. It provides further details on the fabrication process of the silicon photomodulator and a description of the experimental setup. It also includes additional reconstruction data and imaging performance evaluation.

\section{Photomodulator}
\label{Sec:suppl:Photomodulator}
Two photomodulators are used in this work, both of which are made by surface passivation of high resistivity (100)-oriented silicon wafers. The wafers undergo a stringent wet chemical clean, which is immediately followed by transfer to the load lock of a Veeco Fiji G2 plasma-enhanced atomic layer deposition (ALD) tool. Both sides of the wafers are passivated with 20 nm Al$_2$O$_3$, deposited by ALD at 200~$\degree$C using trimethylaluminium (TMA) and O$_2$ plasma as precursors. Following the Al$_2$O$_3$ deposition, the wafers are annealed in a quartz tube furnace for 30 min in air at 450~$\degree$C which increases the photogenerated charge carrier lifetime and modulation efficiency \cite{hooper2019high}. Further details of the ALD process are described in Ref. \cite{grant2024activation}. 

\begin{figure*}[htbp]
\centering
\fbox{\includegraphics[width=0.45\linewidth]{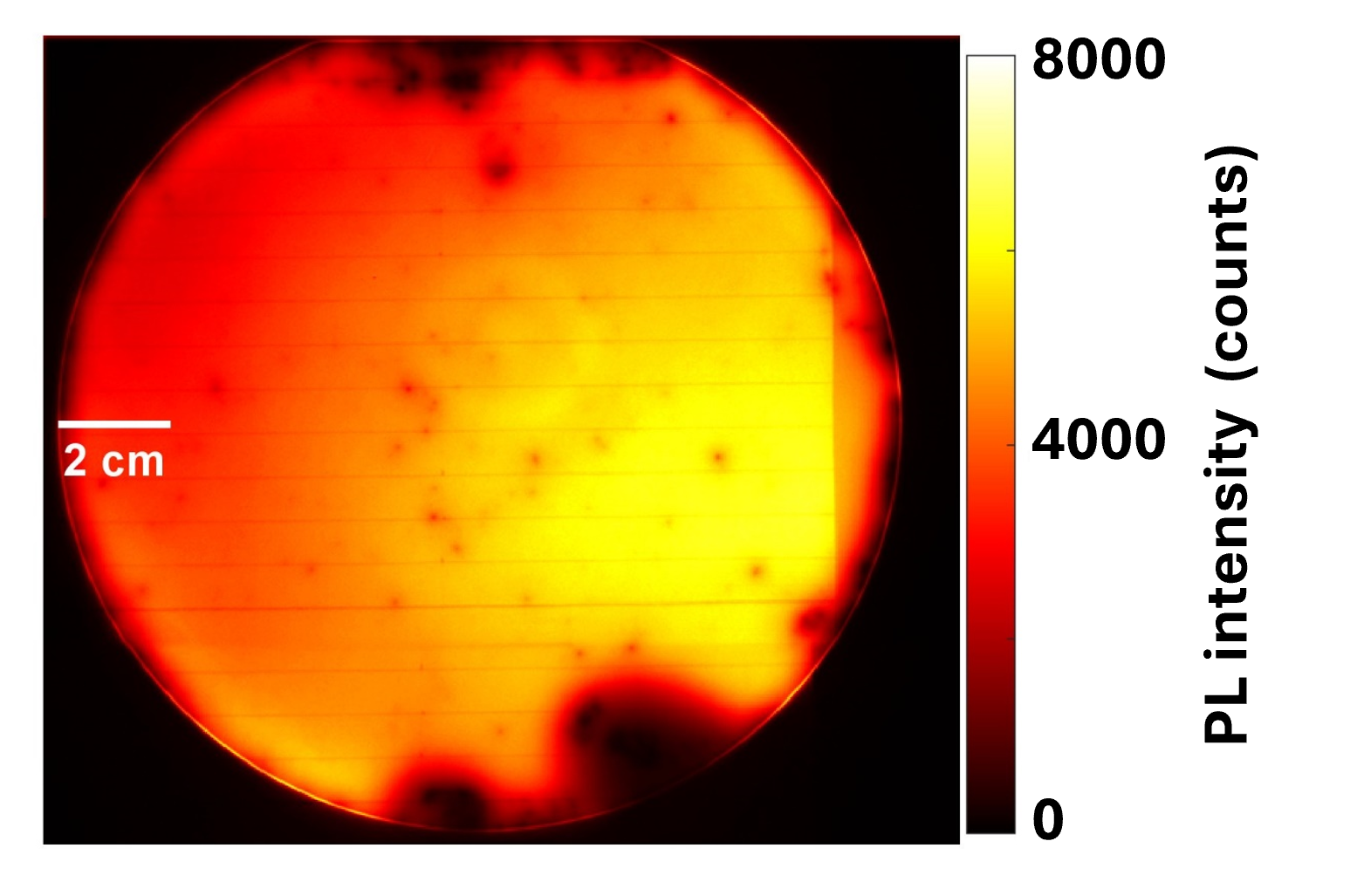}}
\caption{Photoluminescence image of the wafer with a thickness of 380~$\upmu$m, showing the varying carrier lifetime throughout the wafer.}
\label{fig:PL}
\end{figure*}

The lifetime uniformity of the 380~$\upmu$m thick wafer is evaluated by photoluminescence (PL) imaging using an LED array (650 nm), and captured using an ANDOR Apogee Alta F32 CCD camera cooled to -20 °C by a chiller. Fig.~\ref{fig:PL} shows the captured PL image of the wafer, demonstrating the spatial variation in carrier lifetime across it. 

\begin{figure*}[htbp]
\centering
\fbox{\includegraphics[width=0.9\linewidth]{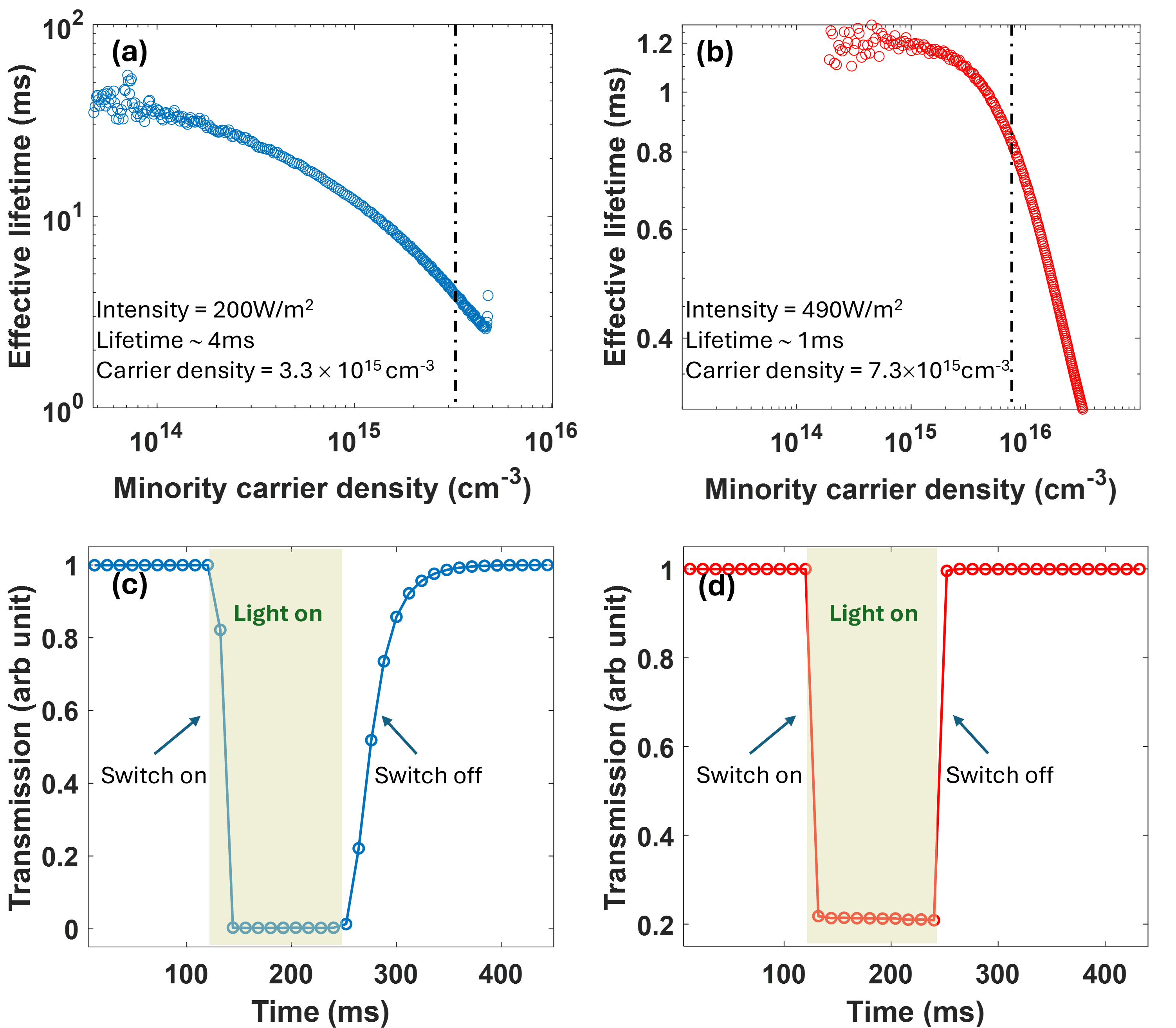}}
\caption{Effective carrier lifetime as a function of the minority carrier density for silicon wafers with thicknesses of (a) 380~$\upmu$m and (b) 100~$\upmu$m, respectively. The dashed vertical lines indicate the corresponding values used in the paper. (c, d) Temporal response of the normalized THz transmission under optical illumination for the 380~$\upmu$m and 100~$\upmu$m wafers, respectively. The shaded area represents the time for which optical illumination was applied.}
\label{fig:lifetime}
\end{figure*}

Both wafers used in this study are fabricated using the same process. However, owing to their differing thicknesses and variations in surface passivation quality, they exhibit different carrier lifetimes as shown in Fig. \ref{fig:lifetime} (a, b). The effective carrier lifetime of each wafer is measured by the photoconductance decay method using Sinton Instruments WCT-120PL lifetime system in transient mode; the given values are averaged over 5 successive measurements. The effective carrier lifetime decreases with increasing carrier density (illumination intensity) which is attributed to enhanced Auger recombination at high injection levels.

The difference in carrier lifetime between wafers directly influences the temporal switching response of the wafers under optical illumination. Fig. \ref{fig:lifetime} (c, d) shows the normalized transmission of the THz beam with time when optical excitation is applied and then removed. As expected, the longer carrier lifetime (380~$\upmu$m thick) wafer in (c) exhibits slower switching times and a deeper modulation than the 100~$\upmu$m thick wafer in (d). Note that for significant modulation depths the switch on and off times diverge from the effective lifetime as shown in (c) \cite{hooper2022engineering}. For this reason the faster 100~$\upmu$m thick modulator is used for the video-rate demonstrations in this work.

\section{Experimental setup}

\begin{figure*}[htbp]
\centering
\fbox{\includegraphics[width=0.9\linewidth]{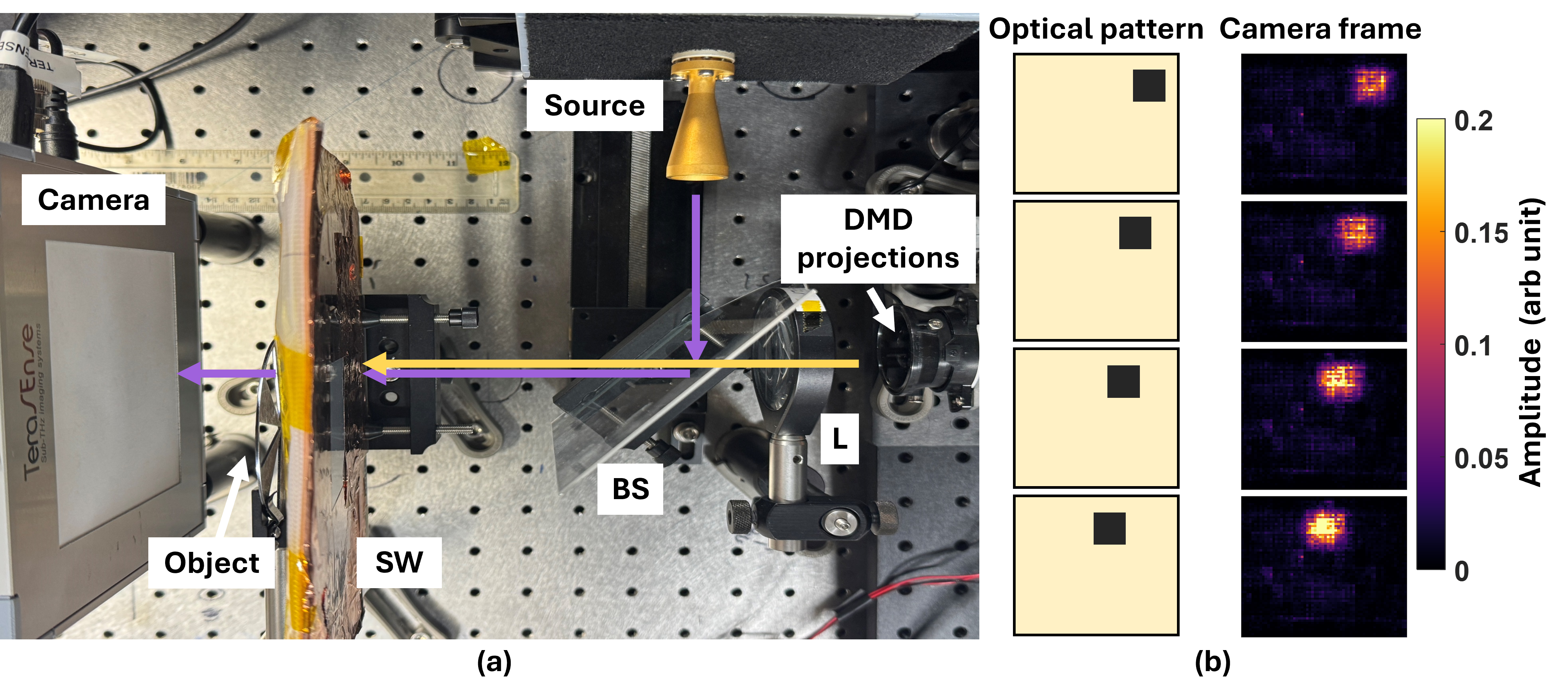}}
\caption{(a) Annotated image of the experimental setup used to record ptychographic imaging data ; SW: silicon wafer, BS: beam splitter, and L: lens. (b) Example of the four optically generated apertures captured by the camera, with corresponding visible light patterns; these frames show the THz beam passing through a non-illuminated region. They were obtained by positioning the camera just behind the silicon wafer and are used as probes for the reconstruction.}
\label{fig:setup}
\end{figure*}

The annotated experimental setup, shown in Fig. \ref{fig:setup} (a), consists of a 95 GHz source and a DMD ($1024~\times~768$ pixels resolution) coupled to a visible light source. The DMD generates structured optical patterns that are transmitted through a beam splitter (BS) using a commercial projection system. An additional plano-convex lens $L$ with a focal length of 175 mm is added $\sim$ 80~mm after the projector aperture for the video rate demonstrations in Fig. 4 of the main text, to increase the illumination intensity. The 95 GHz beam is reflected by the BS and combined with the structured visible light at the silicon wafer. Sub-regions of the DMD are used to define the effective field of view, corresponding to 75~mm $\times$ 75~mm for static imaging and 50~mm $\times$ 50~mm for video measurements, respectively. Each optical pattern contains dark (unexposed) regions, forming a localized aperture for the 95 GHz beam to illuminate the object. The object is scanned by sequentially updating these optical patterns. The camera is placed 50 mm away from the object to capture the resulting diffraction patterns required for the reconstruction. Camera acquisition triggers the DMD to project the next pattern, which is controlled via the LabVIEW environment.

Fig. \ref{fig:setup} (b) illustrates the four example patterns out of a total of 64 (arranged in an 8 × 8 grid) used for raster scanning of the sample. Here, the silicon wafer is fully illuminated with optical light except for a small square region, and the camera is placed directly behind the silicon wafer without any object. The right column of the figure shows the transmission of the THz beam which is used to define the probe function.

\section{Additional reconstruction results}
This section provides information supporting the results shown in Fig. 2 in the main text.
\subsection{Unnormalized images}
\begin{figure*}[htbp]
\centering
\fbox{\includegraphics[width=.9\linewidth]{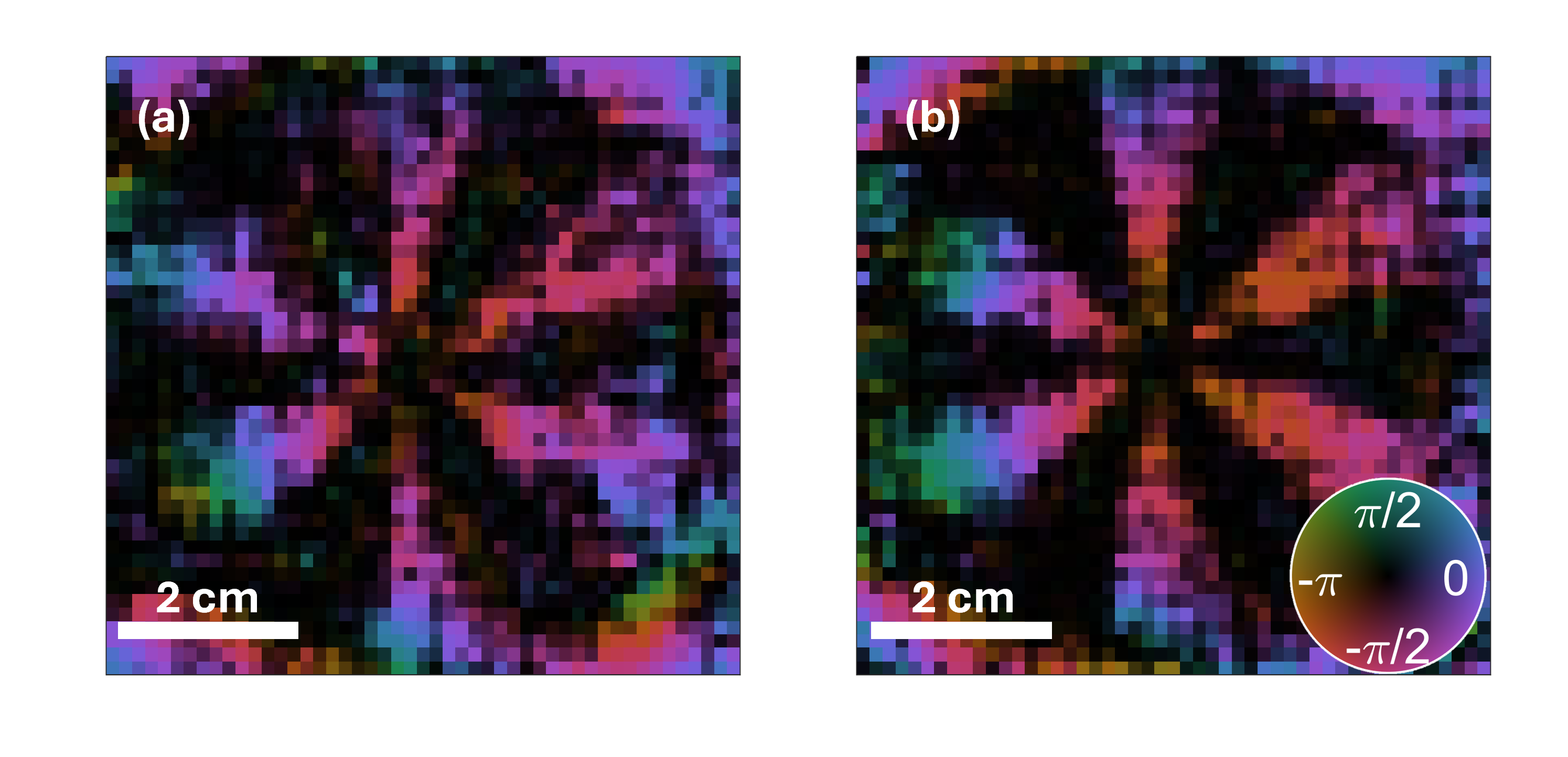}}
\caption{Unnormalized reconstructions of the Siemens star acquired using (a) raster and (b) Hadamard bases, the quadratic color variation shows the phase curvature of the incident THz beam.}
\label{fig:unnormalised_figures}
\end{figure*}

\begin{figure*}[htbp]
\centering
\fbox{\includegraphics[width=0.9\linewidth]{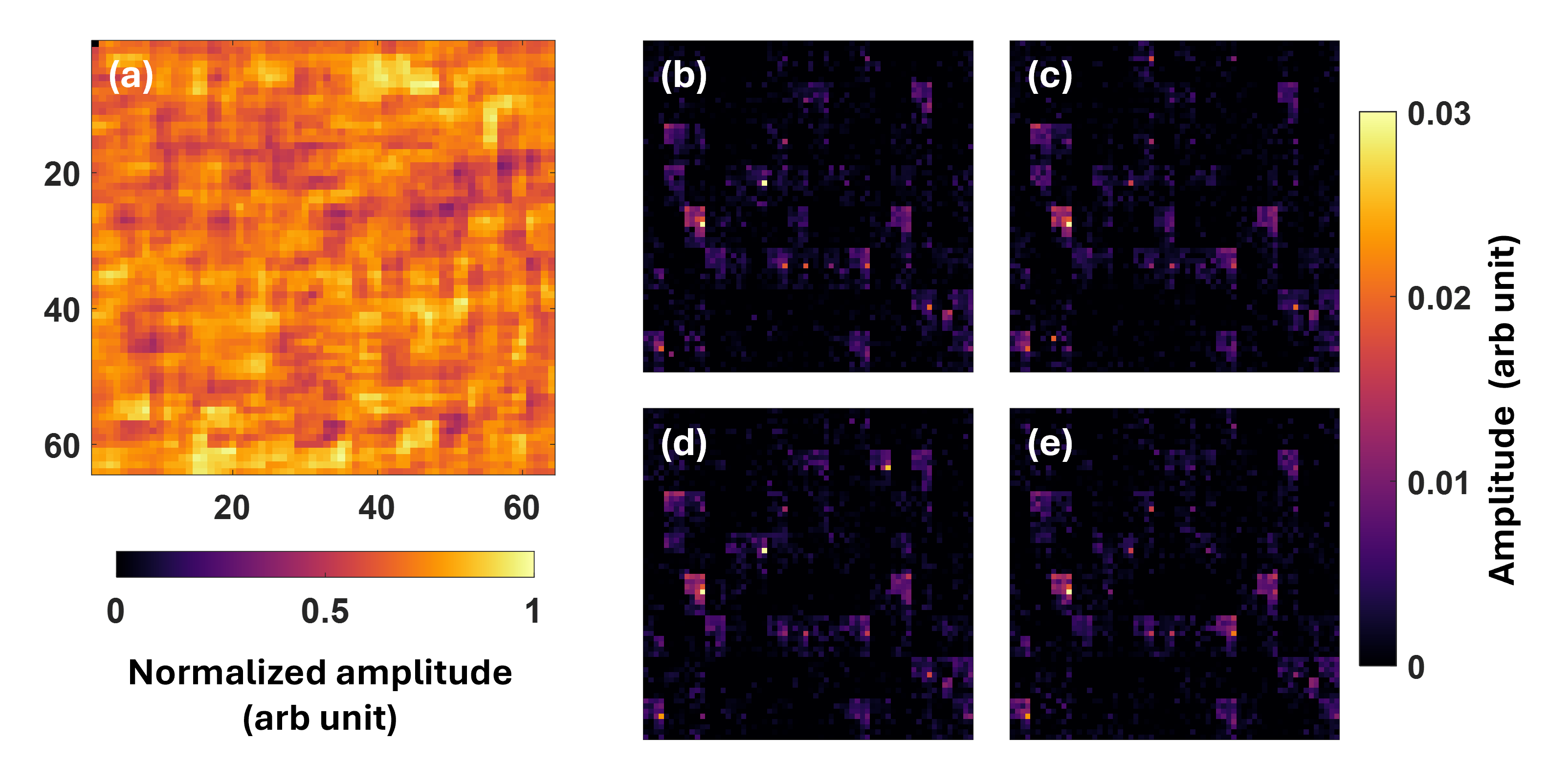}}
\caption{(a) Spatially varying response of the camera pixels measured under uniform illumination. (b–e) Four camera frames acquired at 45 fps with no incident beam, showing fluctuating pixel response across the sensor area.}
\label{fig:camerapixels}
\end{figure*}

The unnormalized images of the Siemens star obtained using raster and Hadamard bases are shown in Fig. \ref{fig:unnormalised_figures}. The image reconstructed with raster scanning exhibits a lower signal-to-noise ratio (SNR), with some measurement noise attributed to the non-uniform and fluctuating pixel response of the camera pixels, as illustrated in Fig. \ref{fig:camerapixels}. In both images, since a uniform probe phase is assumed during reconstruction, the phase curvature of the incident beam is mapped onto the reconstructed object. This leads to the quadratic background color variation in these unnormalised images.

\subsection{Spatial-resolution estimation}
\begin{figure*}[htbp]
\centering
\fbox{\includegraphics[width=0.9\linewidth]{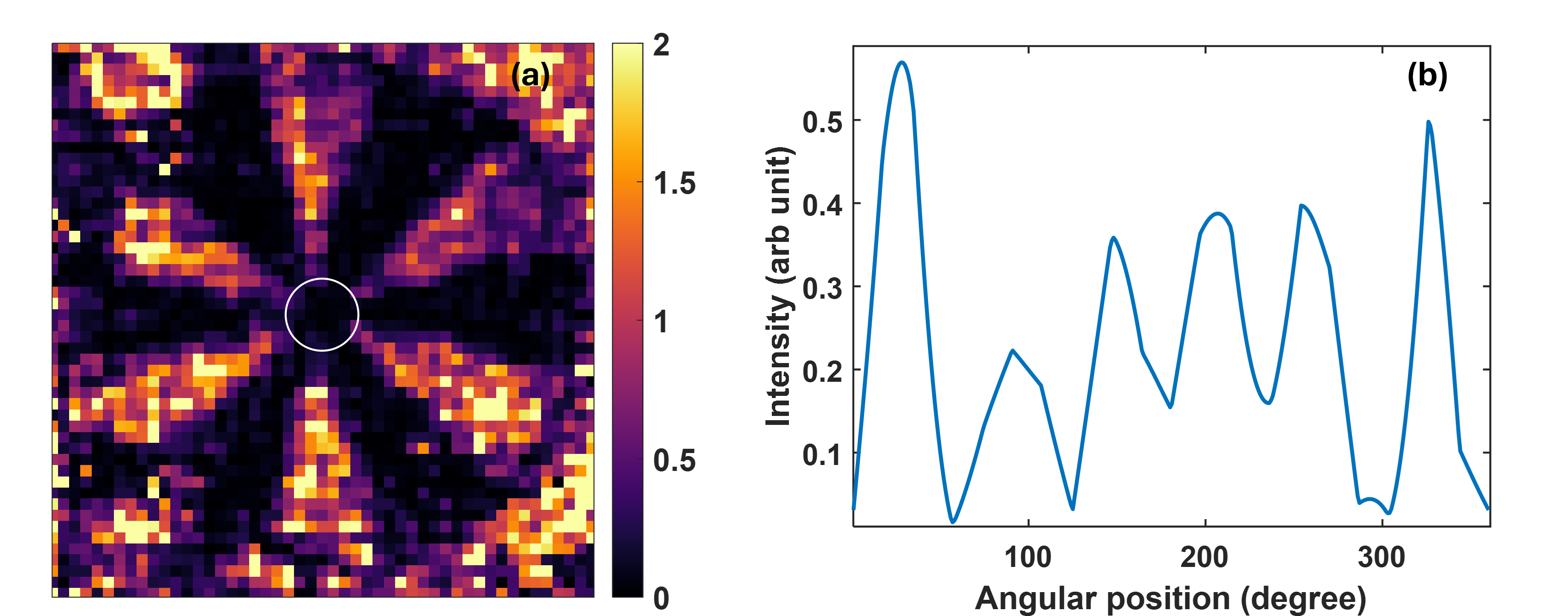}}
\caption{(a) Reconstructed intensity image of the Siemens star generated using the Hadamard basis. The white circle shows a representative concentric circle. (b) Intensity profile along the angular position of the white circle.}
\label{fig:resolution}
\end{figure*}

The spatial resolution is determined by analyzing intensity profiles extracted along the multiple concentric circles centered on the Siemens star. A representative concentric circle is shown in Fig. \ref{fig:resolution} (a), together with the corresponding angular intensity profile in \ref{fig:resolution} (b). For each adjacent spoke pair, concentric circles are drawn, and the minimum radius at which the spokes could be resolved as separate intensity peaks is identified. The corresponding spatial resolution is then calculated from this radius. The reported spatial resolution (2.1$\pm$0.3~mm) is the average of the values obtained for all adjacent spoke pairs.

\subsection{Error prediction}
\begin{figure*}[htbp]
\centering
\fbox{\includegraphics[width=0.45\linewidth]{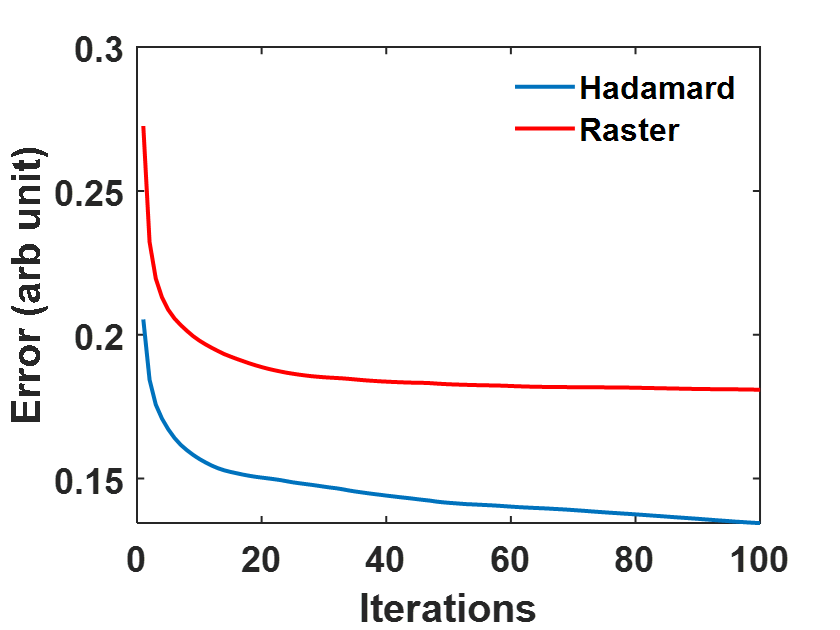}}
\caption{The convergence of reconstruction error  with number of iterations for raster and Hadamard bases. }
\label{fig:convergence}
\end{figure*}

The reconstruction error is quantified using the structural similarity index (SSIM) \cite{wang2004image} between the reconstructed and measured intensity images at each iteration, and is defined as (1 - SSIM). Fig. \ref{fig:convergence} compares the error behavior of reconstructions performed using raster and Hadamard basis projections. The Hadamard illumination-based reconstruction exhibits faster convergence and achieves a lower error as compared to raster scanning. In this case, each measurement simultaneously records the diffraction information from multiple regions of the object,  resulting in faster convergence and a lower reconstruction error.

\subsection{Refractive index calculation}
\begin{figure*}[htbp]
\centering
\fbox{\includegraphics[width=0.95\linewidth]{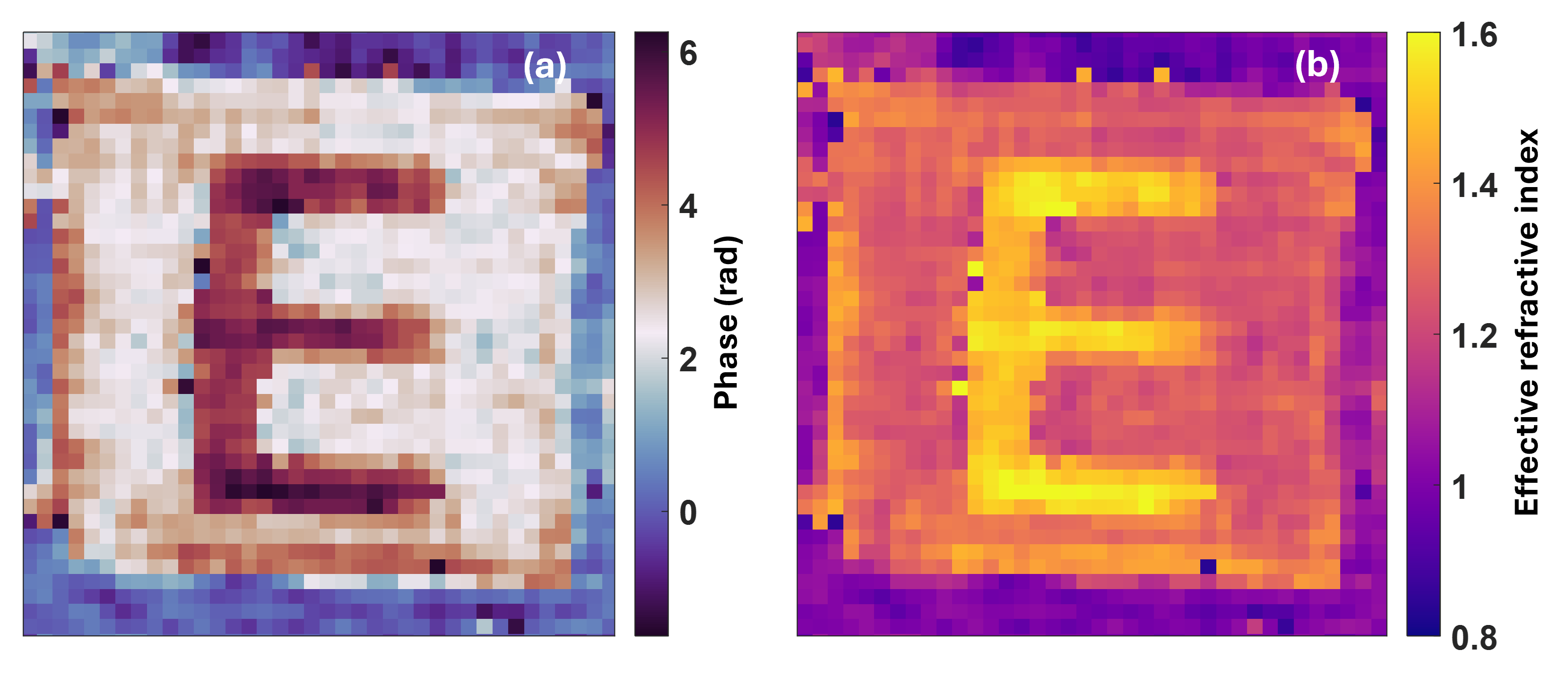}}
\caption{(a) Phase image of the 3D-printed dielectric object after constrained phase unwrapping, (b) calculated effective refractive index; the colormaps are from Ref. \cite{matplotlib_colormaps_matlab}.}
\label{fig:quant_phase}
\end{figure*}

The reconstructed phase map of the 3D-printed dielectric object was unwrapped using a constrained phase unwrapping, guided by prior knowledge of the sample thickness (5~mm) and a constraint on the expected refractive index range \cite{price2020subterahertz}. Since ptychographic phase retrieval yields relative phase rather than absolute values, the known refractive index of the image background (air) is used to provide a corrective global phase offset. The effective refractive index, shown in Fig. \ref{fig:quant_phase} (b),  is calculated from the unwrapped phase, Fig. \ref{fig:quant_phase} (a), using the following relation $n = 1 + \frac{\phi \lambda}{2\pi t}$ \cite{joseph2026decoupling}, where t is the thickness of the block and $\phi$ is the measured unwrapped phase. From the index distribution map, we clearly see a difference in the refractive index of the block and inclusion letter E, which ranges from an index of $\sim$1.2 for 20\% infill density to approximately the homogeneous index$\sim$1.6 at 100\% infill density \cite{price2020subterahertz}. 

\section{Measurement parameters}
Table \ref{tab:Details} summarizes the acquisition parameters used to collect the ptychographic datasets for each figure provided in the paper. Here the number of projections refers to the total number of optical patterns used to reconstruct a single image of the object. Similarly, the total acquisition time represents the time required to collect all frames needed for the reconstruction of one image. For data acquisition, we capture a certain number of camera frames for a given optical projection, this is defined by the camera frames per projection (fpp) in the table. Note that only the final camera frame acquired for each projection is used for reconstructions shown in Fig. 2 and 3. In contrast, all acquired camera frames are used for the video reconstruction shown in Fig. 4.

\begin{table*}[htbp]
\caption{Details of the acquisition parameters for each figure; fpp is camera frames per projection.}
  \label{tab:Details}
  \centering
\begin{tabular}{cccccc}
\hline
Figure & No. of projections & Camera fpp & Camera frame rate (fps) & Total acquisition time (s) \\
\hline
2a & 64 (Raster) & 12 & 45  & 17 \\
2(b-d) & 64 (Hadamard) & 12 & 45  & 17 \\
3a & 64 (Random) & 12 & 45  & 17 \\
3b & 32 (Random) & 12 & 45  & 8.5\\
3c & 8 (Random) & 12 & 45  & 2.1 \\
3d & 4 (Random) & 12 & 45  & 1 \\
4 & 2 (Random) & 2 & 67  & 0.06 \\
\hline
\end{tabular}
\end{table*}

\section{Video-rate imaging}
Supplementary Visualization 1, referenced in the main text shows a reference optical video of the dynamic object, along with the corresponding amplitude and phase reconstructions. At the default 30 fps playback speed this corresponds to the real-time motion of the fluid droplets. This dataset has been processed with a $3~\times~3~\times~3$ median filter, spanning both transverse spatial and temporal dimensions. This improves the visual clarity of the still images used for Fig. 4 of the main text, at the cost of a reduction in effective spatial and temporal bandwidths. For completeness, in Supplementary Visualization 2 we show the unfiltered dataset, in which the motion of the droplets may still be clearly observed. 
The optical camera data is recorded off-axis ($\sim 45^{\circ}$) from the THz camera due to geometrical constraints. To partially compensate for this, the optical video data undergoes an affine transformation consisting primarily of a horizontal scaling by a factor of $\sqrt{2}$ together with a small horizontal shear. This transformation approximately corrects for the difference in viewing geometry between the optical and THz cameras, although some residual discrepancies remain. Note also that due to the viewing angle the left-hand side of the plastic container in Fig. 4 (a) of the main text contains refracted images of the colored droplets and has therefore been rendered partially transparent to increase visualization clarity.
\bibliography{sample}
\end{document}